\documentclass[12pt]{article}
\usepackage{graphicx}
\usepackage{amsmath}
\usepackage{amssymb}
\usepackage{caption2}
\begin{document}
\bibliographystyle{prsty}
\begin{center}
{\large {\bf \sc{  Possible tetraquark  states   in the $\pi^+ \chi_{c1}$ invariant mass distribution }}} \\[2mm]
Zhi-Gang Wang \footnote{E-mail,wangzgyiti@yahoo.com.cn.  }     \\
 Department of Physics, North China Electric Power University,
Baoding 071003, P. R. China \footnote{Mailing address.} \\
Kavli Institute for Theoretical Physics China, Institute of
Theoretical Physics, Chinese Academy of sciences, Beijing 100080, P.
R. China

\end{center}

\begin{abstract}
In this article, we assume that  there   exist hidden charmed
tetraquark states with the spin-parity $J^P=1^-$, and calculate
their masses with  the QCD sum rules.  The numerical result
indicates that the masses of the vector hidden charmed tetraquark
states are about   $M_{Z}=(5.12\pm0.15)\,\rm{GeV}$ or
$M_{Z}=(5.16\pm0.16)\,\rm{GeV}$, which are
   inconsistent with the experimental data
    on  the $\pi^+ \chi_{c1}$ invariant mass distribution.
    The hidden charmed mesons $Z_1$,
$Z_2$ or $Z$ may be scalar hidden charmed tetraquark states,
hadro-charmonium resonances or  molecular states.
\end{abstract}



\section{Introduction}

The Babar, Belle, CLEO, D0, CDF and FOCUS collaborations have
discovered (or confirmed) a large number of charmonium-like states
\cite{Belle-x3940,Belle-chi2c2,Belle-x3872,CDF-x3872,D0-x3872,BaBar-x3872,
BaBar-y4260,Belle-y4008,Belle-y3940,BaBar-y4325,Belle-y4660},
  and revitalized  the interest  in the
spectroscopy of the charmonium states
\cite{review1,review2,review3,review4}.

The $X(3940)$ decaying into $D^*\bar{D}$ and the $Z(3930)$  decaying
into $D\bar{D}$ have been tentatively  identified as candidates for
the missing charmonium states $\eta_c''$  and $\chi_{c2}'$
 respectively. The $X(3872)$ decaying into
$\pi^+\pi^- J/\psi$, $\pi^+\pi^-\pi^0 J/\psi$, the $Y(4260)$,
$Y(4008)$ decaying into $\pi^+\pi^- J/\psi$, the $Y(3940)$ decaying
into $\omega J/\psi$, the $Y(4325)$, $Y(4360)$, $Y(4660)$
 decaying into $\pi^+\pi^-\psi'$
 have odd properties comparing with the expectations of the charmonium models \cite{review1,review2,review3,review4}.

Many possible  assignments for those states have been suggested,
such as multiquark states (whether the molecular type
\cite{molecular1,molecular2} or the diquark-antidiquark type
\cite{tetraquark1,tetraquark2,tetraquark3}), hybrid states
\cite{hybrid1,hybrid2,hybrid3}, charmonium states modified by nearby
thresholds \cite{cc-thresh2,cc-thresh3},  threshold cusps
\cite{cusp1}, etc.

The  observed decay channels  are  $J/\psi \pi^+ \pi^-$
 or $\psi'\pi^+\pi^-$, an essential ingredient
 for understanding the structures of those
mesons is whether or not  the $\pi\pi$ comes from a resonance state.
For example,  there is an indication that  the $Y(4660)$ has a well
defined intermediate state in  the $\pi\pi$ invariant mass
distribution, which is consistent with the scalar meson $f_0(980)$
\cite{BaBar-conf}, the $Y(4660)$ can be taken as a $f_0(980) \psi'$
bound state \cite{FKGuo08}; though other interpretations such as a
baryonium state \cite{Qiao07} or a canonical  $5^3S_1$ $c\bar{c}$
state \cite{DingYan} are not excluded.

 The $Z^+(4430)$  observed in the $  \psi^\prime\pi^+$ decay mode  is the most
interesting subject \cite{Belle-z4430}.   It can't be a pure
$c\bar{c}$ state due to the positive charge. There are  many
theoretical interpretations for its structures, such as the
hadro-charmonium resonance \cite{review3,Voloshin0803}, the
molecular $D^*D_1(D_1')$ state
\cite{Meng07,Lee07-sum,Liu07,Ding07,Braaten07,Liu0803}, the
tetraquark state \cite{Rosner07,Maiani07,Gershtein07,Li07,Liu08},
the cusp in the $D^*D_1$ channel \cite{Bugg07},  the radially
excited state of the $D_s$ \cite{Matsuki0805}, etc. We can
distinguish the multiquark states
 from the hybrids or charmonia with the criterion of
non-zero charge.

Recently the Belle collaboration  reported the first observation of
two resonance-like structures (thereafter we will denote them as
$Z_1$ and $Z_2$ respectively) in the $\pi^+\chi_{c1}$ invariant mass
distribution near $4.1 \,\rm{GeV}$ in the exclusive $\bar{B}^0\to
K^- \pi^+ \chi_{c1}$ decays \cite{Belle-chipi}. Their quark contents
must be some special combinations of the $c\bar{c} u\bar{d}$, just
like the $Z^+(4430)$, they cannot be the conventional mesons. The
Breit-Wigner
 masses and widths are about
$M_1=4051\pm14^{+20}_{-41} \,\rm{MeV}$,
$\Gamma_1=82^{+21}_{-17}$$^{+47}_{-22}\, \rm{MeV}$,
$M_2=4248^{+44}_{-29}$$^{+180}_{-35}\,\rm{MeV}$ and
$\Gamma_2=177^{+54}_{-39}$$^{+316}_{-61}\,\rm{MeV}$.
 The significance of each of the $\pi^+\chi_{c1}$ structures
exceeds $5 \sigma$, including the effects of systematics from
various fit models.

The $Z$  (denote the $Z_1$ and $Z_2$)  lie about
$(0.5-0.6)\,\rm{GeV}$ above the $\pi^+\chi_{c1}$ threshold, the
decay $ Z \to \pi^+\chi_{c1}$ can take place with the "fall-apart"
mechanism and it is OZI super-allowed, which can take into account
the large width naturally. The spins of the $Z_1$ and $Z_2$ are not
determined yet, they can be scalar or vector states
\cite{Belle-chipi}.

If they are scalar mesons, the decays $Z \to \pi^+\chi_{c1}$ occur
through the relative $P$-wave with the phenomenological lagrangian
$\mathcal {L}=g\chi^\alpha ( \pi \partial_\alpha Z-Z\partial_\alpha
\pi)$. On the other hand, if they are vector mesons, the decays
occur through the relative $S$-wave with the phenomenological
lagrangian $\mathcal {L}=g\chi^\alpha Z_\alpha \pi $.

The typical decay mode $Z \to D^+ \bar{D}^0$ is  kinematically
allowed, and the width may be comparable with the corresponding ones
of the decay  mode $Z \to \pi^+ \chi_{c1}$, we can determine the
spins of the $Z$ with the angular distributions of the final states
$D^+\bar{D}^0$. If the decays  $Z \to D^+ \bar{D}^0$ are not
observed (or the widths  are rather narrow), the $Z$ may be
hadro-charmonium resonances \cite{review3}, i.e.  bound states of a
relatively compact charmonium ($\chi_{c1}$) inside a light hadron
($\pi^+$) having a larger spatial size, the decays $ Z \to
\pi^+\chi_{c1}$ occur with the "fall-apart" mechanism;   the decays
$Z \to D^+ \bar{D}^0$ take place through the final-state
re-scattering effects, $Z \to \pi^+ \chi_{c1} \to D^+ \bar{D}^0$,
the widths may be very narrow.

The masses of the $D^+$ and $\bar{D}^0$ are about
$M_D=1.87\,\rm{GeV}$, the $Z$ may also be  $P$-wave $D^+ \bar{D}^0$
molecular states with the spin-parity $1^-$, as the additional
contribution from the relative $P$-wave is about $0.5 \, \rm{GeV}$
in the potential  quark models, the decays $Z \to D^+ \bar{D}^0   $
occur through "fall-apart"  mechanism and have much larger widths
than the corresponding  decays
  $Z \to D^+ \bar{D}^0 \to\pi^+ \chi_{c1} $, which occur through
  final-state re-scattering effects.   One may also think that they
  are $D^*D_1$ (or $D^*D'_1$) molecular states,
$M_{D^*}=2.01\, \rm{GeV}$,  $M_{D_1}=2.42\, \rm{GeV}$,
  $M_{D'_1}=2.43\, \rm{GeV}$, the bound energy is about $(0.2-0.4)
  \,\rm{GeV}$, which may be beyond capacity of the one-$\pi$ and
  one-$\sigma$  exchange.

The mass is a fundamental parameter in describing a hadron, in order
to identify  the $Z_1$ and $Z_2$ as tetraquark states, we must prove
that the masses of the corresponding tetraquark states lie in the
region $(4.1-4.3)\, \rm{GeV}$. Furthermore, whether or not there
exist such hidden tetraquark configurations is of great importance
itself, because it provides a new opportunity for a deeper
understanding of low energy QCD.

In this article, we assume  that the hidden  charmed mesons $Z_1$
and $Z_2$ are vector tetraquark states, which  consist of a
pseudoscalar (scalar) diquark   and an axial-vector (vector)
antidiquark, and study their masses with the QCD sum rules
\cite{SVZ79,Reinders85}. As their spins are not determined yet, they
may also be scalar hidden charmed tetraquark states, we  study this
possibility with the QCD sum rules \cite{Wang08072}.

In the QCD sum rules, operator product expansion is used to expand
the time-ordered currents into a series of quark and gluon
condensates which parameterize the long distance properties of the
QCD vacuum. Based on current-hadron duality, we can obtain copious
information about the hadronic parameters at the phenomenological
side.

 It is difficult to distinguish the mesons $Z_1$ and $Z_2$ with the QCD sum
 rules approach,  as the mass gap between them is very
 small.  The mesons $Z_1$ and $Z_2$ lie in the region $(4.0-4.3) \, \rm{GeV}$, we study whether
or not  there exist $1^-$ hidden charmed tetraquark states $Z$ in
this energy region and possible tetraquark identification
  of the mesons $Z_1$ and $Z_2$.  Furthermore,
  the $\pi^+ \chi_{c1}$ invariant mass distribution amplitude can
  also be represented by a single Breit-Wigner mass formula,   $M_Z=4150^{+31}_{-16}\,
  \rm{MeV}$ and $\Gamma_Z=352^{+99}_{-43}\, \rm{MeV}$ \cite{Belle-chipi}.

The article is arranged as follows:  we derive the QCD sum rules for
the masses and pole residues  of  the $Z$  in section 2; in section
3, numerical results and discussions; section 4 is reserved for
conclusion.

\section{QCD sum rules for  the tetraquark states  $Z$ }
In the following, we write down  the two-point correlation functions
$\Pi_{\mu\nu}(p)$ (denote $\Pi^{J}_{\mu\nu}(p)$ and
$\Pi^{\eta}_{\mu\nu}(p)$) in the QCD sum rules,
\begin{eqnarray}
\Pi^{J/\eta}_{\mu\nu}(p)&=&i\int d^4x e^{ip \cdot x} \langle
0|T\left\{J/\eta_\mu(x)J/\eta^{\dagger}_\nu(0)\right\}|0\rangle \, ,  \\
J_\mu(x)&=& \epsilon^{ijk}\epsilon^{imn}u_j^T(x) C c_k(x)
\bar{c}_m(x) \gamma_\mu C \bar{d}_n^T(x)\, , \\
\eta_\mu(x)&=& \epsilon^{ijk}\epsilon^{imn}u_j^T(x) C\gamma_5 c_k(x)
\bar{c}_m(x) \gamma_5\gamma_\mu C  \bar{d}_n^T(x)\, , \\
f_{Z}M_{Z}^4\epsilon_\mu  &=& \langle 0|J/\eta_\mu(0)|Z(p)\rangle\,
,
\end{eqnarray}
we choose  the vector currents $J_\mu(x)$ ($C-C\gamma_\mu$ type) and
$\eta_\mu(x)$ ($C\gamma_5-C\gamma_\mu\gamma_5$ type) to interpolate
the tetraquark states $Z$, the $f_Z$ is the pole residue and the
$\epsilon_\mu$ is the polarization  vector. If there exist  vector
tetraquark states $Z$ in the $\pi^+ \chi_{c1}$ invariant mass
distribution, it is convenient to construct the tetraquark currents
with the pseudoscalar (scalar) diquark  and axial-vector (vector)
antidiquark, as the mesons $\pi$ and $\chi_{c1}$ have the
spin-parity $0^-$ and $1^+$ respectively, the decays $Z \to \pi^+
\chi_{c1}$ can occur through relative $S$-wave. We can also
interpolate the vector tetraquark states with the currents
$J^1_\mu(x)$ and $\eta^1_\mu(x)$,
\begin{eqnarray}
J^1_\mu(x)&=& \epsilon^{ijk}\epsilon^{imn}u_j^T(x) C\gamma_\mu
c_k(x) \bar{c}_m(x) C  \bar{d}_n^T(x) \, ,\\
\eta^1_\mu(x)&=& \epsilon^{ijk}\epsilon^{imn}u_j^T(x) C\gamma_\mu
\gamma_5 c_k(x) \bar{c}_m(x) \gamma_5 C  \bar{d}_n^T(x) \, ,
\end{eqnarray}
which consist of a pseudoscalar (scalar) antidiquark  and an
axial-vector (vector) diquark. Our analytical results indicate that
the current $J_\mu(x)$ ($\eta_\mu(x)$) and $J^1_\mu(x)$
($\eta^1_\mu(x)$) lead to the same expression. The special
superpositions  $tJ_\mu(x)+(1-t)J^1_\mu(x)$ and
$t\eta_\mu(x)+(1-t)\eta^1_\mu(x)$  cannot  improve the predictions
remarkably, where $t=0-1$.

The correlation functions $\Pi_{\mu\nu}(p)$ can be decomposed as
follows,
\begin{eqnarray}
\Pi_{\mu\nu}(p)&=&(-g_{\mu\nu}+\frac{p_\mu
p_\nu}{p^2})\Pi_1(p^2)+\frac{p_\mu p_\nu}{p^2} \Pi_0(p^2) +\cdots,
\end{eqnarray}
due to  Lorentz covariance. The invariant functions $\Pi_1$ and
$\Pi_0$ stand for the contributions from the vector and scalar
mesons, respectively. In this article, we choose the tensor
structure $g_{\mu\nu}-\frac{p_\mu p_\nu}{p^2}$ to study the masses
of the vector mesons.

Basing on the quark-hadron duality \cite{SVZ79,Reinders85}, we can
insert  a complete series of intermediate states with the same
quantum numbers as the current operators $J_\mu(x)$ and
$\eta_\mu(x)$ into the correlation functions $\Pi_{\mu\nu}(p)$  to
obtain the hadronic representation. After isolating the ground state
contribution from the pole terms of the $Z$, we get the following
result,
\begin{eqnarray}
\Pi_{\mu\nu}(p)&=&\frac{f_{Z}^2M_{Z}^8}{M_{Z}^2-p^2}\left[-g_{\mu\nu}+\frac{p_\mu
p_\nu}{p^2} \right] +\cdots \, \, .
\end{eqnarray}

In the following, we briefly outline  operator product expansion for
the correlation functions $\Pi_{\mu\nu }(p)$  in perturbative QCD
theory. The calculations are performed at   large space-like
momentum region $p^2\ll 0$. We write down the "full" propagators
$S_{ij}(x)$ and $C_{ij}(x)$ of a massive quark in the presence of
the vacuum condensates firstly \cite{Reinders85},
\begin{eqnarray}
S_{ij}(x)&=& \frac{i\delta_{ij}\!\not\!{x}}{ 2\pi^2x^4}
-\frac{\delta_{ij}m_q}{4\pi^2x^2}-\frac{\delta_{ij}}{12}\langle
\bar{q}q\rangle +\frac{i\delta_{ij}}{48}m_q
\langle\bar{q}q\rangle\!\not\!{x}-\nonumber\\
&&\frac{\delta_{ij}x^2}{192}\langle \bar{q}g_s\sigma Gq\rangle
 +\frac{i\delta_{ij}x^2}{1152}m_q\langle \bar{q}g_s\sigma
 Gq\rangle \!\not\!{x}-\nonumber\\
&&\frac{i}{32\pi^2x^2} G^{ij}_{\mu\nu} (\!\not\!{x}
\sigma^{\mu\nu}+\sigma^{\mu\nu} \!\not\!{x})  +\cdots \, ,
\end{eqnarray}
\begin{eqnarray}
C_{ij}(x)&=&\frac{i}{(2\pi)^4}\int d^4k e^{-ik \cdot x} \left\{
\frac{\delta_{ij}}{\!\not\!{k}-m_c}
-\frac{g_sG^{\alpha\beta}_{ij}}{4}\frac{\sigma_{\alpha\beta}(\!\not\!{k}+m_c)+(\!\not\!{k}+m_c)\sigma_{\alpha\beta}}{(k^2-m_c^2)^2}\right.\nonumber\\
&&\left.+\frac{\pi^2}{3} \langle \frac{\alpha_sGG}{\pi}\rangle
\delta_{ij}m_c \frac{k^2+m_c\!\not\!{k}}{(k^2-m_c^2)^4}
+\cdots\right\} \, ,
\end{eqnarray}
where $\langle \bar{s}g_s\sigma Gs\rangle=\langle
\bar{s}g_s\sigma_{\alpha\beta} G^{\alpha\beta}s\rangle$  and
$\langle \frac{\alpha_sGG}{\pi}\rangle=\langle
\frac{\alpha_sG_{\alpha\beta}G^{\alpha\beta}}{\pi}\rangle$, then
contract the quark fields in the correlation functions
$\Pi_{\mu\nu}(p)$ with Wick theorem, and obtain the result:
\begin{eqnarray}
\Pi^J_{\mu\nu}(p)&=&i\epsilon^{ijk}\epsilon^{imn}\epsilon^{i'j'k'}\epsilon^{i'm'n'}\int
d^4x e^{ip \cdot x} Tr\left[ C S_{kk'}^T(x) C C_{jj'}(x)
\right]\nonumber\\
&&Tr\left[\gamma_\mu C S_{n'n}^T(-x) C \gamma_\nu C_{m'm}(-x)
\right]\, , \\
\Pi^\eta_{\mu\nu}(p)&=&i\epsilon^{ijk}\epsilon^{imn}\epsilon^{i'j'k'}\epsilon^{i'm'n'}\int
d^4x e^{ip \cdot x} Tr\left[ \gamma_5 C S_{kk'}^T(x) C\gamma_5
C_{jj'}(x)
\right]\nonumber\\
&&Tr\left[\gamma_5\gamma_\mu C S_{n'n}^T(-x) C \gamma_\nu \gamma_5
C_{m'm}(-x) \right]\, .
\end{eqnarray}
Substitute the full $u$, $d$ and $c$ quark propagators into the
correlation functions  $\Pi_{\mu\nu}(p)$ and complete  the integral
in the coordinate space, then integrate over the variables in the
momentum space, we can obtain the correlation functions $\Pi_1(p^2)$
at the level of quark-gluon degrees  of freedom. Once analytical
results are obtained,   then we can take  current-hadron duality
below the threshold $s_0$ and perform  Borel transform  with respect
to the variable $P^2=-p^2$, finally we obtain  the following sum
rules:
\begin{eqnarray}
f_Z^2 M_Z^8 e^{-\frac{M_Z^2}{M^2}}= \int_{4m_c^2}^{s_0} ds
\rho_{J/\eta}(s)e^{-\frac{s}{M^2}} \, ,
\end{eqnarray}
\begin{eqnarray}
\rho_{J/\eta}(s)&=&\frac{1}{3072 \pi^6}
\int_{\alpha_{min}}^{\alpha_{max}}d\alpha
\int_{\beta_{min}}^{1-\alpha} d\beta
\alpha\beta(1-\alpha-\beta)^3(s-\widetilde{m}^2_c)^2(7s-\widetilde{m}^2_c)(5s-3\widetilde{m}^2_c)
\nonumber \\
&&\mp\frac{m_c\langle \bar{q}q\rangle}{32 \pi^4}
\int_{\alpha_{min}}^{\alpha_{max}}d\alpha
\int_{\beta_{min}}^{1-\alpha} d\beta
(1-\alpha-\beta)(s-\widetilde{m}^2_c)\nonumber\\
&&\left[
s(4\beta-3\alpha)+\widetilde{m}^2_c(\alpha-2\beta)\right] \nonumber\\
&& \mp\frac{m_c\langle \bar{q}g_s\sigma Gq\rangle}{64 \pi^4}
\int_{\alpha_{min}}^{\alpha_{max}}d\alpha
\int_{\beta_{min}}^{1-\alpha} d\beta \left[
s(2\alpha-3\beta)-\widetilde{m}^2_c(\alpha-2\beta) \right] \nonumber\\
&&-\frac{m_c^2\langle \bar{q}q\rangle^2}{12 \pi^2}
\int_{\alpha_{min}}^{\alpha_{max}} d\alpha
 \, ,
\end{eqnarray}
where $\alpha_{max}=\frac{1+\sqrt{1-\frac{4m_c^2}{s}}}{2}$,
$\alpha_{min}=\frac{1-\sqrt{1-\frac{4m_c^2}{s}}}{2}$,
$\beta_{min}=\frac{\alpha m_c^2}{\alpha s -m_c^2}$ and
$\widetilde{m}_c^2=\frac{(\alpha+\beta)m_c^2}{\alpha\beta}$.

 We carry out  operator
product expansion to the vacuum condensates adding up to
dimension-6. In calculation, we
 take  assumption of vacuum saturation for  high
dimension vacuum condensates, they  are always
 factorized to lower condensates with vacuum saturation in the QCD sum rules,
  factorization works well in  large $N_c$ limit.
In this article, we take into account the contributions from the
quark condensates,  mixed condensates, and neglect the contributions
from the gluon condensate. In calculation, we observe the
contributions  from the gluon condensate are suppressed by large
denominators and would not play any significant roles
\cite{Wang1,Wang2,Wang3,Wang4,Wang5}. Furthermore, we  neglect the
terms proportional to the $m_q$ ($=m_u=m_d$), their contributions
are of minor importance and can be neglected safely.

 Differentiating  the Eq.(13) with respect to  $\frac{1}{M^2}$, then eliminate the
 pole residue $f_{Z}$, we can obtain two sum rules  for
 the masses  of the $Z$,
 \begin{eqnarray}
 M_Z^2=\frac{ \int_{4m_c^2}^{s_0} ds
\rho_{J/\eta}(s)se^{-\frac{s}{M^2}} }{\int_{4m_c^2}^{s_0} ds
\rho_{J/\eta}(s)e^{-\frac{s}{M^2}}}\, .
\end{eqnarray}

\section{Numerical results and discussions}

The $c$-quark mass appearing in the perturbative terms (see e.g.
Eq.(14)) is usually taken to be the pole mass in the QCD sum rules,
while the choice of the $m_c$ in the leading-order coefficients of
the higher-dimensional terms is arbitrary \cite{Kho9801}. It is
convenient to take the pole mass $m_c=(1.3\pm0.1)\,\rm{GeV}$
\cite{Kho9801}. The $\overline{MS}$ mass $m_c(m_c^2)$ relates with
the pole mass $\hat{m}$ through the relation  $m_c(m_c^2)
=\hat{m}\left[1+\frac{C_F
\alpha_s(m_c^2)}{\pi}+(K-2C_F)\left(\frac{\alpha_s}{\pi}\right)^2+\cdots\right]^{-1}$,
where $K$ depends on the flavor number $n_f$. In this article, we
take the approximation $m_c\approx\hat{m}$ without the $\alpha_s$
corrections for consistency. The value listed in the PDG is
$m_c(m_c^2)=1.27^{+0.07}_{-0.11} \, \rm{GeV}$ \cite{PDG}, it is
reasonable to take the value $m_c=(1.3\pm0.1)\,\rm{GeV}$ in
Ref.\cite{Kho9801}, we also present the result with larger
uncertainty. The vacuum condensates are scale dependent, the average
virtuality of the quarks is characterized  by the Borel parameter
$M^2$, it
 makes sense  to choose $\mu^2=\mathcal {O}(M^2)$. In this article, the
energy scale is taken as $\mu=2\,\rm{GeV}$, $\langle \bar{q}q
\rangle=-(0.26\pm 0.01\, \rm{GeV})^3$,  $\langle \bar{q}g_s\sigma G
q \rangle=m_0^2\langle \bar{q}q \rangle$, $m_0^2=(0.8 \pm
0.2)\,\rm{GeV}^2$ \cite{SVZ79,Reinders85,Ioffe2005}.

In the conventional QCD sum rules \cite{SVZ79,Reinders85}, there are
two criteria (pole dominance and convergence of the operator product
expansion) for choosing  the Borel parameter $M^2$ and threshold
parameter $s_0$. In calculation, we usually  consult the
experimental data in choosing those parameters.

The Belle collaboration  observed the resonance-like structures
$Z_1$ and $Z_2$ in the $\pi^+\chi_{c1}$ invariant mass distribution
near $4.1 \,\rm{GeV}$ in the exclusive $\bar{B}^0\to K^- \pi^+
\chi_{c1}$ decays \cite{Belle-chipi}.  The Breit-Wigner
 masses and the widths are about
$M_1=4051\pm14^{+20}_{-41} \,\rm{MeV}$,
$\Gamma_1=82^{+21}_{-17}$$^{+47}_{-22}\, \rm{MeV}$,
$M_2=4248^{+44}_{-29}$$^{+180}_{-35}\,\rm{MeV}$ and
$\Gamma_2=177^{+54}_{-39}$$^{+316}_{-61}\,\rm{MeV}$. If they are
vector hidden charmed tetraquark states, the central value of the
threshold parameter can be tentatively taken as $s_0=(4.248+0.5)^2\,
\rm{GeV}^2\approx 23 \, \rm{GeV}^2$, where we choose the separation
between the ground states and first radial excited states to be
$0.5\,\rm{GeV}$.

The present experimental knowledge about the phenomenological
hadronic spectral densities of the tetraquark states is  rather
vague, whether or not there exist   tetraquark states is not
confirmed with confidence, and no knowledge about  the high
resonances; we can borrow some ideas from the baryon spectra
\cite{PDG}.

For the octet baryons with the quantum numbers
$I(J^{P})=\frac{1}{2}({\frac{1}{2}}^+)$, the mass of the proton (the
ground state)  is $M_p=938\,\rm{MeV}$, and the mass of the first
radial excited state $N(1440)$ (the Roper resonance) is
$M_{1440}=(1420-1470)\,\rm{MeV}\approx 1440\,\rm{MeV}$ \cite{PDG}.
For the decuplet  baryons with the quantum numbers
$I(J^{P})=\frac{3}{2}({\frac{3}{2}}^+)$ , the mass of  the
$\Delta(1232)$ (the ground state) is
$M_{1232}=(1231-1233)\,\rm{MeV}\approx 1232\,\rm{MeV}$,  and the
mass of the first radial excited state $\Delta(1600)$ is
$M_{1600}=(1550-1700)\,\rm{MeV}\approx 1600\,\rm{MeV}$ \cite{PDG}.
The mass gap between the ground states and first radial excited
states can be chosen as $0.5\, \rm{GeV}$. In this article, the
central value of the threshold parameter  $s_0=23\, \rm{GeV}^2$
makes sense.

However, the threshold parameter  $s_0=23\, \rm{GeV}^2$ cannot
result in a reasonable Borel window, we have to postpone it
tentatively to larger values. It  is  not an indication that
non-existence  of the vector hidden charmed  tetraquark states below
$4.7\, \rm{GeV}$; in other words, the QCD sum rules alone cannot
indicate (non-) existence of the multiquark states strictly.

If the multiquark states exist  indeed, we can release the criterion
of pole dominance and take a more phenomenological analysis with the
QCD sum rules. One may refuse the value extracted from continuum
dominating QCD sum rules as quantitatively reliable if one insists
on that the contribution from the pole term should be larger than
(or about) $50\%$ (for detailed discussions about this subject, one
can consult Ref.\cite{Wang1}). In the present case, the numerical
results indicate that the threshold parameter $s_0 > 30 \,
\rm{GeV}^2$ can lead to possible Borel window, we  take the value
$s_0=(32\pm 1)\, \rm{GeV}^2$.

With the central values of the input parameters, the contribution of
the condensate (of the largest dimension) $\langle
\bar{q}q\rangle^2$ is less than $28\%$ ($32\%$) at the value $M^2
\geq 3.4 \, \rm{GeV}^2 $,  and the contribution decreases quickly to
about $10\%$ ($11\%$) at the value $M^2 = 4.5 \, \rm{GeV}^2 $ for
the $C-C\gamma_\mu$ ($C\gamma_5-C\gamma_\mu\gamma_5$) type
   interpolating current. In this article, the Borel parameter can be  taken
as $M^2 \ge 3.4 \, \rm{GeV}^2 $, we expect the operator product
expansion is convergent.

The contribution of the pole term  is lager  than $50\%$ ($49\%$) at
the value $M^2 \leq 4.5 \, \rm{GeV}^2 $, and the pole contribution
is about $(50-74)\%$ ($(49-71)\%$) at the value $M^2=(3.4-4.5)
\,\rm{GeV}^2$ for the $C-C\gamma_\mu$
($C\gamma_5-C\gamma_\mu\gamma_5$) type
   interpolating current,
again we take the central values of the input parameters.  If we
take into account the uncertainty of the threshold parameter,
$s_0=(32\pm1)\, \rm{GeV}^2$, the pole contribution is about
$(48-77)\%$ ($(45-75)\%$).  The Borel parameter can be taken  as
$M^2=(3.4-4.5) \, \rm{GeV}^2 $, where two criteria of the QCD sum
rules are full filled \cite{SVZ79,Reinders85}. From the Figs.1-2, we
can see that the sum rules are not stable enough below the value
$M^2=3.8\, \rm{GeV}^2$. In the article, the Borel parameter and the
threshold parameter are taken as $M^2=(3.8-4.5) \, \rm{GeV}^2 $ and
$s_0=(32\pm1)\, \rm{GeV}^2$, respectively.

 For the tetraquark states  consist of light flavors, if the perturbative
terms have the main contribution (in the conventional QCD sum rules,
the perturbative terms are always have the main contribution), we
can approximate the spectral density with the perturbative term
(where the $A$ are some numerical coefficients) \cite{Wang0708},
\begin{eqnarray}
B_M\Pi \sim A \int_0^\infty s^4 e^{-\frac{s}{M^2}}ds=A
M^{10}\int_0^\infty t^4 e^{-t}dt \, ,
\end{eqnarray}
take the pole dominance condition,
\begin{eqnarray}
\frac{\int_0^{t_0} t^4 e^{-t}dt}{\int_0^\infty t^4 e^{-t}dt}\geq
50\% \, ,
\end{eqnarray}
and obtain the approximated  relation,
\begin{eqnarray}
t_0&=&\frac{s_0}{M^2}\geq 4.7 \, .
\end{eqnarray}
The superpositions of different interpolating currents can only
change the contributions from different terms in the operator
product expansion, and  improve convergence, they cannot change the
leading behavior of the spectral density $\rho(s)\propto s^4 $ of
the perturbative term \cite{Wang0708}.

This relation is difficult to satisfy for the light flavor
tetraquark states \cite{Wang1,Wang2,Wang3,Wang4,Wang5}, however, it
is not an indication that  non-existence  of the tetraquark states.
The hidden charmed and bottomed tetraquark states, and open bottomed
tetraquark states may satisfy the relation, as they always have
larger Borel parameter $M^2$ and threshold parameter $s_0$
\cite{tetraquark3,Lee07-sum,Narison07,SR4430}. Although the relation
is derived for the light flavor quarks in the massless limit, the
$c$ and $b$ are heavy quarks.

For examples, in Ref.\cite{Narison07}, the authors take the
$X(3872)$ as hidden charmed tetraquark state and calculate its mass
with the QCD sum rules, the Borel parameter and threshold parameter
are taken as $M^2=(2.0-2.8)\,\rm{GeV^2}$ and
$s_0=(17-18)\,\rm{GeV}^2$; in Ref.\cite{Lee07-sum}, the authors take
the $Z(4430)$ as hidden charmed molecular  state and calculate its
mass with $M^2=(2.5-3.1)\,\rm{GeV^2}$ and $s_0=(23-25)\,\rm{GeV}^2$.
In those sum rules,  the relation in Eq.(18) can be well satisfied.

In this article, $s_0/M^2>6.8$, the relation in Eq.(18) is certainly
satisfied. The relation can serve as an additional constraint in
choosing the Borel parameter and threshold parameter, it  alone
cannot lead to satisfactory results, as a number of values of the
Borel parameter and threshold parameter satisfy the relation.

Taking into account all uncertainties of the input parameters,
finally we obtain the values of the masses and pole residues of
 the   $Z$, which are
shown in Figs.(1-2),
\begin{eqnarray}
M_{Z}&=&(5.12\pm0.15)\,\rm{GeV} \, , \nonumber\\
f_{Z}&=& (1.31 \pm0.26 )\times 10^{-4}\,\rm{GeV} \, ;\\
M_{Z}&=&(5.16\pm0.16)\,\rm{GeV} \, , \nonumber\\
f_{Z}&=& (1.25 \pm 0.25)\times 10^{-4}\,\rm{GeV} \, ,
\end{eqnarray}
for the $C-C\gamma_\mu$ type    and
  $C\gamma_5-C\gamma_\mu\gamma_5$ type interpolating currents respectively.
  In numerical calculations, we observe the uncertainties come from
  the parameter $m_0^2$ are very small, while the uncertainties come from
  the parameters $m_c$, $\langle \bar{q}q\rangle$ and $s_0$ are
  comparable with  each other.

  If we take larger uncertainty for the pole mass, $m_c=(1.3\pm 0.2) \,
  \rm{GeV}$, the values of the mass  change to
  $M_{Z}=(5.12\pm0.28)\,\rm{GeV}$ and
  $M_{Z}=(5.16\pm0.30)\,\rm{GeV}$ respectively. Furthermore,  we
  vary  the energy scale from  $\mu^2=m_c^2$ to $\mu^2=4.5\,
  \rm{GeV}^2$, the central value of the mass $M_Z$ changes slowly, less than  $0.05 \,
  \rm{GeV}$.

  The value $\sqrt{s_0}-M_Z\approx 0.5\, \rm{GeV}$ happens to be the energy
  gap  between the ground states and first radial excited
states of the light baryons \cite{PDG}, the threshold parameter
$s_0=(32\pm1)\, \rm{GeV}^2$ makes sense.

At the energy scale $\mu=2\, \rm{GeV}$, $\frac{\alpha_s}{\pi}\approx
0.09$ \cite{PDG}, if the perturbative $\mathcal {O}(\alpha_s)$
corrections to the perturbative term are companied with large
numerical factors, $1+\xi(s,m_c)\frac{\alpha_s}{\pi}$, for example,
$\xi(s,m_c) >\frac{\pi}{\alpha_s}\approx 10$, the contributions may
be large. We can make a crude estimation by multiplying the
perturbative term  with a numerical factor, say
$1+\xi(s,m_c)\frac{\alpha_s}{\pi}=2$, the mass $M_Z$ decreases
slightly, about $0.1\, \rm{GeV}$, the  pole residue changes
remarkably. The main contribution comes from the perturbative term,
the large corrections in the numerator and denominator  cancel out
with each other (see Eq.(15)). In fact, the $\xi(s,m_c)$ are
complicated functions of the energy $s$ and the mass $m_c$, such a
crude estimation may underestimate the  $\mathcal {O}(\alpha_s)$
corrections, the uncertainties originate from the $\mathcal
{O}(\alpha_s)$ corrections maybe larger.

  The hidden charmed tetraquark state with the
spin-parity $1^-$ lie in the region $(5.0-5.3) \, \rm{GeV}$, the
 mesons $Z_1$, $Z_2$ or $Z$ (about $(4.1-4.3) \,
\rm{GeV}$) in the $\pi^+ \chi_{c1}$ invariant mass distribution
cannot be vector tetraquark states. The spins of the $Z$ are not
determined yet, they may be scalar hidden charmed states which  may
lie in the region $(4.1-4.3) \, \rm{GeV}$ \cite{Wang08072}, or more
likely, they are hadro-charmonium resonances \cite{review3},
$P$-wave $D^+\bar{D}^0$ molecular states, or $D_1^+\bar{D}^0+
D^+\bar{D}_1^0$ molecular states; more experimental data are still
needed to identify them.

\begin{figure}
 \centering
 \includegraphics[totalheight=6cm,width=7cm]{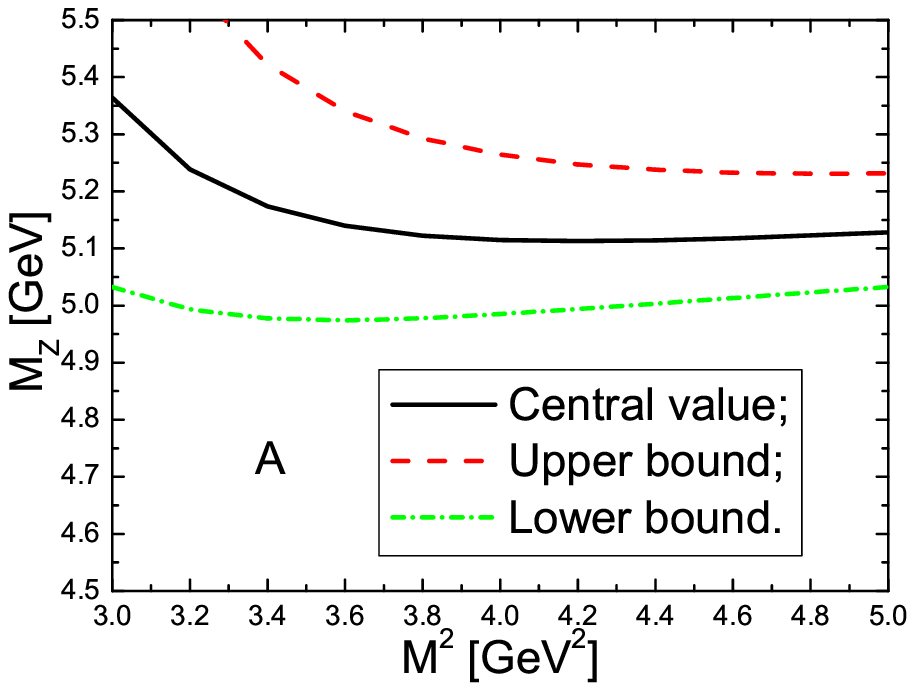}
 \includegraphics[totalheight=6cm,width=7cm]{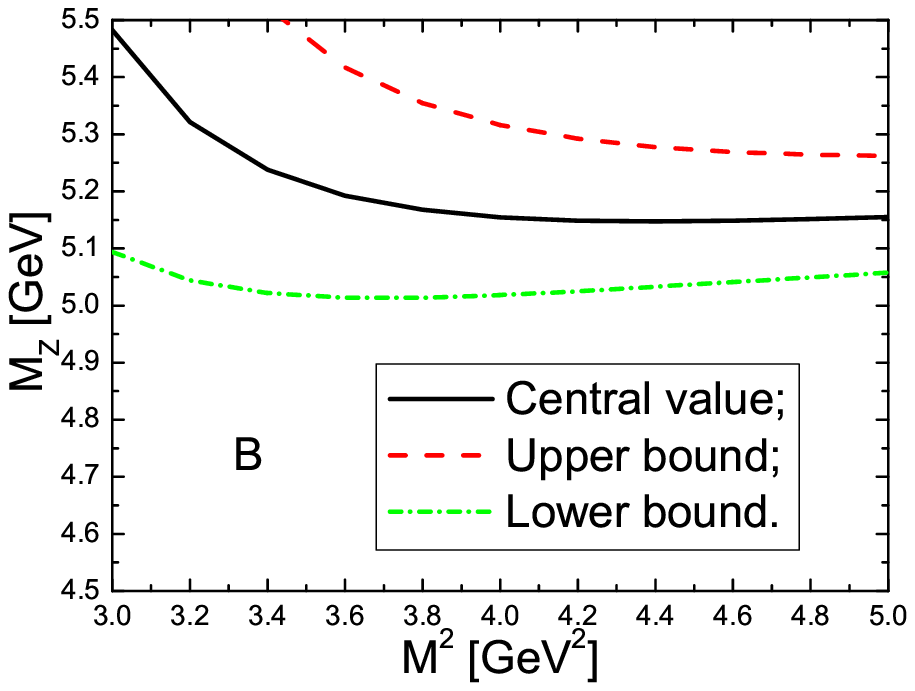}
  \caption{ The mass $M_{Z}$ with variation of the   Borel parameter $M^2$, $A$ for the $C-C\gamma_\mu$ type current
 and $B$ for $C\gamma_5-C\gamma_\mu\gamma_5$ type
   current. }
\end{figure}

\begin{figure}
 \centering
 \includegraphics[totalheight=6cm,width=7cm]{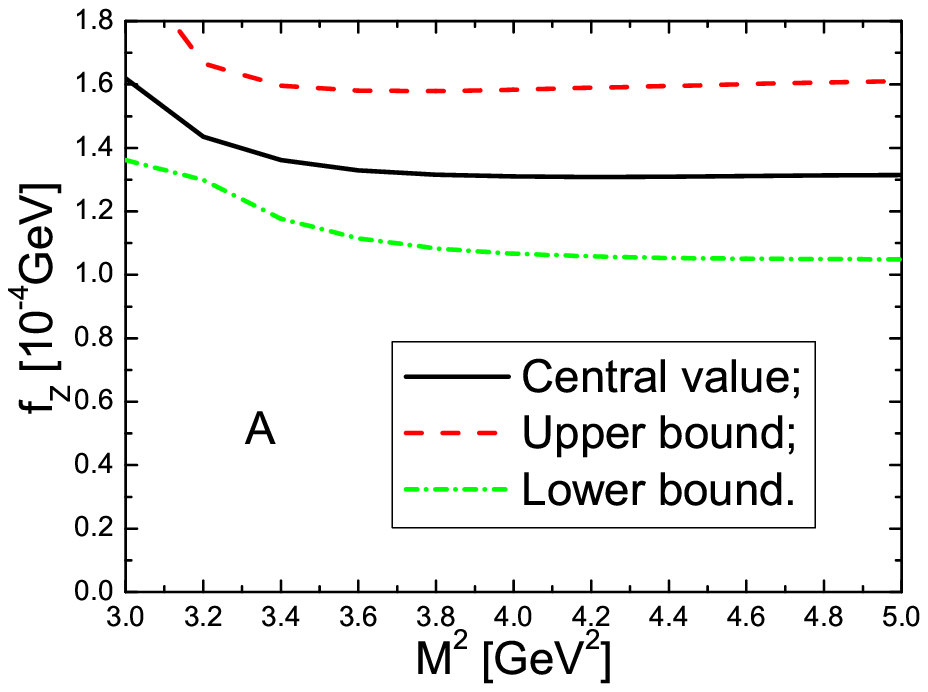}
 \includegraphics[totalheight=6cm,width=7cm]{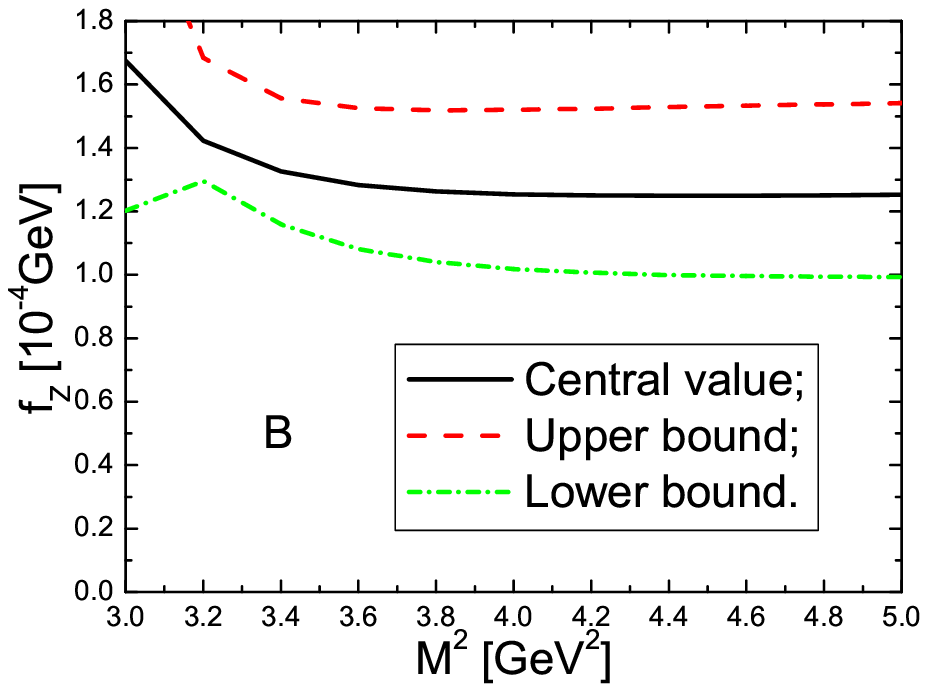}
  \caption{The pole residue  $f_Z$ with variation of the  Borel parameter $M^2$, $A$ for the $C-C\gamma_\mu$ type current
 and $B$ for $C\gamma_5-C\gamma_\mu\gamma_5$ type
   current. }
\end{figure}

\section{Conclusion}
In this article, we assume that  there   exist   hidden charmed
tetraquark states with with the spin-parity $J^P=1^-$, and calculate
their masses with   the QCD sum rules.  The numerical result
  indicates that the masses of the vector hidden charmed tetraquark states are about
  $M_{Z}=(5.12\pm0.15)\,\rm{GeV}$ or $M_{Z}=(5.16\pm0.16)\,\rm{GeV}$, which are
   inconsistent with the experimental data
    on  the $\pi^+ \chi_{c1}$ invariant mass distribution.
    The hidden charmed mesons $Z_1$,
$Z_2$ or $Z$ may be scalar hidden charmed states,  hadro-charmonium
resonances or  molecular states.

\section*{Acknowledgments}
This  work is supported by National Natural Science Foundation,
Grant Number 10775051, and Program for New Century Excellent Talents
in University, Grant Number NCET-07-0282.

\end{document}